\documentclass[aps,amsmath,amssymb,nofootinbib]{revtex4} 
\usepackage{graphicx} 
\usepackage{amsmath}
\usepackage{amsfonts,amsbsy}
\usepackage{amssymb}

\def\be{\begin{equation}}
\def\ee{\end{equation}}
\def\bea{\begin{eqnarray}}
\def\eea{\end{eqnarray}}

\begin{document}

\title{\bf Azimuthal asymmetries and the emergence of ``collectivity''
from multi-particle correlations in high-energy pA collisions}

\author{Adrian Dumitru}
\email{Adrian.Dumitru@baruch.cuny.edu}
\affiliation{Department of Natural Sciences, Baruch College, CUNY,
17 Lexington Avenue, New York, NY 10010, USA}
\affiliation{The Graduate School and University Center, The City University of New York, 365 Fifth Avenue, New York, NY 10016, USA}

\author{Larry McLerran } 
\email{McLerran@bnl.gov}
\affiliation{RIKEN BNL, Brookhaven National Laboratory, 
Upton, NY 11973}
\affiliation{
Physics Department, Brookhaven National Laboratory,
Upton, NY 11973, USA}
\affiliation{
Physics Department, China Central Normal University, Wuhan, China}

\author{Vladimir Skokov}
\email{Vladimir.Skokov@wmich.edu}
\affiliation{Department of Physics, Western Michigan University, Kalamazoo, MI 49008, USA}

\begin{abstract}
We show how angular asymmetries $\sim\cos 2\phi$ can arise in dipole
scattering at high energies. We illustrate the effects due to
anisotropic fluctuations of the saturation momentum of the target with
a finite correlation length in the transverse impact parameter plane,
i.e.\ from a domain-like structure. We compute the two-particle
azimuthal cumulant in this model including both one-particle
factorizable as well as genuine two-particle non-factorizable
contributions to the two-particle cross section. We also compute the
full BBGKY hierarchy for the four-particle azimuthal cumulant and find
that only the fully factorizable contribution to $c_2\{4\}$ is
negative while all contributions from genuine two, three and
four-particle correlations are positive. Our results may provide some
qualitative insight into the origin of azimuthal asymmetries in p+Pb
collisions at the LHC which reveal a change of sign of $c_2\{4\}$ in
high-multiplicity events.
\end{abstract}

\maketitle

\section{Introduction}

Large azimuthal asymmetries have been observed in p+Pb collisions at
the LHC~\cite{pPb_ALICE,pPb_ALICE2,pPb_ATLAS,pPb_CMS} and in d+Au
collisions at RHIC~\cite{dAu_RHIC}. These asymmetries are usually
measured via multi-particle angular correlations (see below) and were
found to extend over a long range in rapidity. Causality then requires
that the correlations originate from the earliest times of the
collision~\cite{Dumitru:2008wn}. Furthermore, the data shows that the
asymmetries persist up to rather high transverse momenta, well beyond
$p_\perp \sim 1$~GeV. Recent data by the ATLAS collaboration, for
example, shows that large ``elliptic'' ($v_2$)
asymmetries in p+Pb collisions at $\surd s=5$~TeV persist up to $p_\perp
= 10$~GeV~\cite{Aad:2014lta}. Therefore, it is important to develop an
understanding of their origin in terms of semi-hard (short distance)
QCD
dynamics~\cite{Kovchegov:2002nf,pAridge,KovnerLublinsky,Kovchegov:2012nd,Gyulassy:2014cfa,McLerran:2014uka,Levin:2014kwa,Ozonder:2014sra}.

The ALICE collaboration has measured the two- and four-particle $v_2$
cumulants in p+Pb collisions at 5~TeV as a function of multiplicity,
see~Figs. 1 and 4 in Ref.~\cite{pPb_ALICE2}. These cumulants are
defined as~\cite{Borghini:2001vi}
\bea
c_2\{2\} &=& \left< \exp \; 2i (\phi_1-\phi_2) \right>~, \\
c_2\{4\} &=& \langle \exp\; 2i (\phi_1 +\phi_2 - \phi_3
-\phi_4)  \rangle -
2 \, \langle \exp\; 2i (\phi_1 -\phi_3) \rangle\; \langle
\exp\; 2i (\phi_2 -\phi_4) \rangle~.
\eea
Here, $\langle\cdot\rangle$ denotes an average over the corresponding
azimuthal angles weighted by the two- or four-particle distribution,
respectively. The two-particle cumulant with a rapidity gap suppresses
contributions from resonance decays and jet fragmentation; it depends
weakly on multiplicity and is positive over the entire range of
multiplicity.  On the other hand the four-particle cumulant,
$c_2\{4\}$, decreases monotonically and changes sign to become
negative in high multiplicity events, an effect also seen by the CMS
collaboration (see second paper in~\cite{pPb_CMS}). As shown below,
this requires an anisotropy of the single-particle angular
distribution. In the soft, long wavelength regime, $c_2\{4\}$ is
negative when hydrodynamic flow dominates over ``non-flow''
correlations~\cite{Avsar:2010rf}. In this paper we perform a first
computation of all connected and disconnected contributions to the
cumulants in the short distance regime using a model that allows
for anisotropic ``domains'' of the color-electric fields $\vec{E}$ of
the target~\cite{KovnerLublinsky,Dumitru:2014dra}.

\section{Calculation}
Our discussion is based on the dipole model of high-energy
interactions~\cite{MuellerDipole}. We consider scattering of a dipole
of size $r\sim 1/p_\perp$ from the target described by a particular
configuration of the (color) electric field $E^i \sim F^{+i}$.  For a
small dipole $\vec r \equiv \vec x - \vec y$ the leading C-even
interaction with the target is given by
\be
\label{eq:S1}
S-1 = \frac{1}{2N_c}\,
\mathrm{tr}\, (ig \, \vec r \cdot \vec E)^2 ~,
\ee
with a C-odd correction at order $(igr)^3$ which is not considered here
because it does not contribute to $\sim\cos\, 2\phi$
asymmetries~\cite{Dumitru:2014dra}. Equation~(\ref{eq:S1}) arises
from an expansion of the S-matrix, $\mathrm{tr}\; V(\vec x)
V^\dagger(\vec y)/N_c$, in powers of $\vec r$, where
\be
V(\vec x) = {\cal P} \exp\left(ig \int dx^- A^+(x^-,\vec x)\right)
\ee
is the path-ordered Wilson line describing the propagation of a charge
in the field of the (right-moving) target. We focus on the S-matrix
for a fundamental charge though the calculation could be repeated for a
charge in the adjoint representation yielding the same results for
$c_2\{2\}$ and $c_2\{4\}$.

To obtain the cross section the scattering matrix is averaged over the
configurations of the $\vec E$ field of the target. Averaging over
{\em all} such configurations leads to
\be \label{eq:Ei_Ej_iso}
\langle S \rangle -1  =  
\frac{(ig)^2}{2N_c} r^i r^j \left<\mathrm{tr}\; E^i(\vec b)E^j(\vec b)
\right> = -
\frac{1}{4} r^2 Q_s^2(\vec b) \log \frac{1}{r\Lambda}~
\ee
in the leading log approximation, $\log 1/r\Lambda \gg 1$. Here, $Q_s(\vec
b)$ denotes the saturation scale below which non-linear effects become
significant. In what follows we shall assume a very large nucleus and drop
the dependence of the average saturation momentum on $\vec b$.

Equation~(\ref{eq:Ei_Ej_iso}) corresponds to the single-particle cross
section averaged over all configurations of $\vec E(\vec b)$ in the
target and is, of course, isotropic. On the other hand, for any
particular configuration the S-matrix does exhibit an angular
dependence, c.f.\ for example Fig.~7 in
Ref.~\cite{Dumitru:2011vk}. The idea that anisotropic fluctuations of
the saturation momentum would induce $v_n\ne0$ has been presented
previously in
Refs.~\cite{KovnerLublinsky,Dumitru:2014dra,Noronha:2014vva}. Hence,
to evaluate the amplitude of the angular modulation of the S-matrix we
perform the average subject to the constraint
\be \label{eq:Ei_Ej_a}
\frac{(ig)^2}{2N_c} r^i r^j \left<\mathrm{tr}\; E^i(\vec b_1) E^j(\vec b_2)
\right>_{\hat a} = -
\frac{1}{4} r^2 Q_s^2 \log \frac{1}{r\Lambda}\, \Delta(\vec
b_1-\vec b_2)
\left(1-{\cal A}+2{\cal A}\, (\hat r \cdot \hat a)^2\right)~.
\ee
That is, we divide the target ensemble into classes such that for a given
class the anisotropic part of the electric field correlator in the
vicinity of $\vec b$ (within a given ``domain'') points in a specific
direction. The summation over all classes, which corresponds to an
integration over the directions $\hat a$, is performed only after the
$m$-particle angular cumulant has been evaluated. The quantity ${\cal
  A}$ in eq.~(\ref{eq:Ei_Ej_a}) is the amplitude of anisotropy of the
electric field correlator.

For simplicity, as we mentioned above, in our current analysis we
singled out only fluctuations of $\hat a$ while possible fluctuations
of $Q_s$ and ${\cal A}$ are averaged out in Eq.~\eqref{eq:Ei_Ej_a}.
The results could be extended to account for fluctuations of $Q_s$
and ${\cal A}$ in the future.

The domain structure of the field is described by the two-point correlation
function
\be \label{eq:<EE>_Domain}
\Delta(\vec b_1-\vec b_2) = 
\exp\left(-\frac{|\vec b_1-\vec b_2|^2}{\xi^2} \right)~,
\ee
where $\xi$ denotes the correlation length. We assume a Gaussian
correlation function, other options do not change our results
qualitatively. To simplify the notation we introduce
\be
\frac{1}{N_D} \equiv \frac{1}{S_\perp^2} \int d^2b_1 d^2b_2 \; \Delta(\vec
b_1-\vec b_2)
= \frac{\pi\, \xi^2}{S_\perp}~,
\ee
which is the area of a domain divided by the area of the collision
zone, in other words, the inverse number of
domains. Equation~(\ref{eq:<EE>_Domain}) essentially describes the
correlations of the saturation momentum $Q_s$ in the transverse plane.

We can now compute the angular distribution for scattering of a single
dipole, for a fixed $\hat a$. Using Eqs.~(\ref{eq:Ei_Ej_a}) and
performing a Fourier transform to momentum space, as well as an
average over the impact parameter, we arrive at
\be \label{eq:dN1_Fourier_normalized}
\left(\frac{1}{\pi}\frac{dN}{dk^2}\right)^{-1}\; \frac{dN}{d^2k} =
1 - 2{\cal A} + 4{\cal A}\, (\hat k \cdot \hat a)^2~.
\ee
Hence, the one-particle $v_2$ cumulant
\be
v_2\{1\} \equiv \left< e^{2i(\phi_k-\phi_a)}\right>_{\hat a} = {\cal A}~.
\ee
To avoid confusion let us stress that here $\langle\cdot\rangle$
refers to a {\em different} average than the average over $\vec
E$-field configurations from above; it is simply an average over the
azimuthal angle $\phi_k$ weighted by the
distribution~(\ref{eq:dN1_Fourier_normalized}).

We now proceed to two-particle distributions. The averages over $\vec
E$-field configurations shall be performed assuming a Gaussian
action~\cite{MV} and a color diagonal four-point function although in
general additional contributions could
appear~\cite{KovnerLublinsky,Dumitru:2010mv}. Then the two-particle
S-matrix for fixed $\hat a$ is given by
\begin{eqnarray}
&&\langle S_2 \rangle -1 =
\left(\frac{(ig)^2}{2N_c}\right)^2 \left<
\mathrm{tr}\; (\vec r_1\cdot \vec E(\vec b_1))^2 \; \mathrm{tr}\; (\vec
r_2\cdot \vec E(\vec b_2))^2 \right>_{\hat a}~ \\
 &=&  \frac{(ig)^4}{4N_c^2} 
\int \frac{d\phi_{a'}}{2\pi}
\left<\mathrm{tr}\; \left(\vec r_1\cdot \vec E(\vec b_1)\right)^2
\right>_{\hat a}
\left<\mathrm{tr}\; \left(\vec r_2\cdot \vec E(\vec b_2) \right)^2
\right>_{\hat a'} C(\hat a, \hat a') \label{eq:S2_disc}\\
& & +  \frac{(ig)^4}{4 N_c^2}  \left\langle
\mathrm{tr}\;
\left( \vec r_1 \cdot \vec E (\vec b_1)  \right)^2
\; \mathrm{tr}\;
\left( \vec r_2 \cdot \vec E (\vec b_2)  \right)^2
\right\rangle^\mathrm{conn.}_{\hat a}~.
\label{Eq:S2}
\end{eqnarray}
The factorizable (disconnected) contribution involves the correlations
of the directions of $\vec E(\vec b)$ in the impact parameter plane;
we employ $C(\hat a, \hat a') = 2\pi\, \delta(\phi_a-\phi_{a'})\,
\Delta(\vec b_1-\vec b_2)$. Averaging over impact parameters gives
\bea
& & 
\frac{(ig)^4}{4N_c^2} \int \frac{d^2b_1}{S_\perp}\frac{d^2b_2}{S_\perp}
\int \frac{d\phi_{a'}}{2\pi}
\left<\mathrm{tr}\; \left(\vec r_1\cdot \vec E(\vec b_1)\right)^2
\right>_{\hat a}
\left<\mathrm{tr}\; \left(\vec r_2\cdot \vec E(\vec b_2) \right)^2
\right>_{\hat a'} C(\hat a, \hat a') \\
&=&
\frac{1}{N_D}\; \frac{1}{16} r_1^2 r_2^2 Q_s^4 \log\frac{1}{r_1\Lambda}
\log\frac{1}{r_2\Lambda}
\left(1 - {\cal A} + 2{\cal A}\, (\hat r_1 \cdot \hat a)^2\right)
\left(1 - {\cal A} + 2{\cal A}\, (\hat r_2 \cdot \hat a)^2\right)\\ 
&=&
\frac{1}{N_D} \frac{dN_1}{\pi dr_1^2}  \frac{dN_2}{\pi dr_2^2}  
\left(1 - {\cal A} + 2{\cal A}\, (\hat r_1 \cdot \hat a)^2\right)
\left(1 - {\cal A} + 2{\cal A}\, (\hat r_2 \cdot \hat a)^2\right)~.
\eea
In this expression the prefactor $1/N_D$ arises due to the fact that
the orientation of the electric field is approximately constant only
over distance scales of order the correlation length $\xi$.
Multiplying the Fourier transform of this expression by
$\exp(2i(\phi_1-\phi_2))$ and averaging over the azimuthal angles
leads to the disconnected (single-particle factorizable) contribution
to $(v_2\{2\})^2$:
\be
\label{Eq:Factc2}
\left< e^{2i(\phi_1-\phi_2)}\right>^\mathrm{disc.}_{\hat a} =
\frac{1}{N_D + \frac{1}{2(N_c^2-1)}(1+{\cal A}^2) }\; (v_2\{1\})^2~.
\ee
Note that this is independent of the global direction $\hat a$ relative to
which we define $\phi_1$ and $\phi_2$ and so the final average over
$\hat a$ is trivial.
The additional term in the denominator originates from the connected contribution to the 
normalization.

The connected contribution from Eq.~(\ref{Eq:S2}) is
\bea
&&\frac{(ig)^4}{4 N_c^2}  \left\langle
\mathrm{tr}\;
\left( \vec r_1 \cdot \vec E (\vec b_1)  \right)^2
\; \mathrm{tr}\;
\left( \vec r_2 \cdot \vec E (\vec b_2)  \right)^2
\right\rangle^\mathrm{conn.}_{\hat a}  = \\
&& \frac18 \frac{r_1^2 r_2^2 Q_s^4}{N_c^2-1} \log\frac{1}{r_1\Lambda}
\log\frac{1}{r_2\Lambda} \; \Delta^2(\vec b_1 -\vec b_2)
\left[ \cos(\phi_1 -  \phi_2) + 2{\cal A}
  \left( 2\cos\left( \phi_1 -\phi_a \right)  \cos\left( \phi_2 -\phi_a
  \right) - \cos(\phi_1-\phi_2)
  \right)      \right]^2~.
\eea
Averaging over impact parameters produces a factor
\be
\frac{1}{S_\perp^2} \int d^2b_1 d^2b_2 \, \Delta^2(\vec b_1 -\vec
  b_2) = \frac{1}{2 N_D}~,
\ee
so that the connected contribution to the two-particle cumulant
becomes
\bea
\left\langle e^{2i (\phi_1-\phi_2) }
\right\rangle^\mathrm{conn}_{\hat a} &\equiv& \left.
\int \frac{d\phi_1}{2\pi}\frac{d\phi_2}{2\pi}\; e^{2i (\phi_1-\phi_2)}
\left[ \frac{d N_2(\hat a)}{d^2k_1 d^2k_2} - 
\frac{d N_1(\hat a)}{d^2k_1 }
\frac{d N_1(\hat a)}{d^2k_2 }   
\right]  \right/ \int \frac{d\phi_1}{2\pi}\frac{d\phi_2}{2\pi}\; 
 \frac{d N_2(\hat a)}{d^2k_1 d^2k_2}  \\
&=&
\frac{1}{ N_D+\frac{1}{2(N_c^2-1)}\left(1+{\cal A}^2\right) } \;
\frac{1}{4(N_c^2-1)}~.  \label{eq:v2conn_result}
\eea
As before, here the average $\langle\cdot\rangle$ on the l.h.s.\ is an
average over $\phi_1$ and $\phi_2$ but does not involve averaging over
$\vec E$-field configurations since the one- and two-particle
distributions have already been averaged over all such configurations
corresponding to a given $\hat a$. However, the r.h.s.\ is independent
of $\hat a$ so that the final average over its direction is
trivial. Also, for ${\cal A}={\cal O}(1/N_c)$ the first factor on the
r.h.s.\ of Eqs.~(\ref{Eq:Factc2},\ref{eq:v2conn_result}) can be
approximated by $1/N_D$ so that in all, $v_2\{2\}$ is then given by
\be \label{eq:v2_2}
(v_2\{2\})^2 \equiv \left\langle e^{2i (\phi_1-\phi_2) } \right\rangle
=\frac{1}{N_D}\;
\left({\cal A}^2 + \frac{1}{4(N_c^2-1)}\right)~.
\ee
The first term is the square of the single-particle $v_2\{1\}$; it is
scaled by $1/N_D$ since both particles have to scatter from the same
domain. The second contribution corresponds to genuine
non-factorizable two-particle correlations. Both contributions are
positive; nonetheless Eq.~(\ref{eq:v2_2}) reveals the existence of two
distinct regimes. For
\be
{\cal A} \gg \frac{1}{N_c}
\ee
the ellipticity is mainly due to the asymmetry of the single-particle
distribution induced by the $\vec E$-field domains. In the opposite limit
\be
{\cal A} \ll \frac{1}{N_c},
\ee
$v_2\{2\}$ is mainly due to genuine two-particle correlations.

Expression~(\ref{eq:v2_2}) applies when {\em both} particles have
sufficiently high transverse momenta as we have approximated both of
their S-matrices by their leading small-$r$ behavior $\sim
\mathrm{tr}\; (\vec r_i\cdot\vec E)^2$. On the other hand,
experimentally one typically considers angular correlations of a hard
with a softer particle. Recent numerical
computations~\cite{Lappi:2015vha} of $c_2\{2\}$ which do not expand
the S-matrices show that hard-soft correlations exhibit a fall-off
with the transverse momentum of the hard particle. This is due to a
decorrelation of the anisotropy axis in a high-$p_T$ bin with that of
the bulk.

The four particle cumulant exhibits qualitatively different behavior
in the regimes of ``small'' vs.\ ``large'' ${\cal A}$. For general
${\cal A}$, $c_2\{4\}$ is given by
\bea
c_2\{4\} &=& \langle \exp\left( 2i (\phi_1 +\phi_2 - \phi_3
-\phi_4)\right)  \rangle -
2 \, \langle \exp\left( 2i (\phi_1 -\phi_3) \right)  \rangle\; \langle
\exp\left( 2i (\phi_2 -\phi_4) \right) \rangle
\label{eq:c4} \\
&=& - \frac{1}{N_D^3}(v_2\{1\})^4 +  \langle \exp\left( 2i (\phi_1
+\phi_2 - \phi_3
-\phi_4)\right)  \rangle^\mathrm{conn.} \\   
&+& \frac{1}{N_D} \langle
\exp \left( 2i (\phi_1+\phi_2) \right)  \rangle ^\mathrm{conn.}\;  
\langle \exp \left( - 2 i (\phi_3+\phi_4) \right)  \rangle
^\mathrm{conn.} + \frac{4}{N_D} v_2\{1\}  
\langle \exp \left(  2 i (\phi_1+\phi_2-\phi_3) \right)  \rangle
^\mathrm{conn.} \label{Eq:c4ani_a}\\  
&+& \frac{1}{N_D^2} (v_2\{1\})^2 \langle \exp \left( - 2 i
(\phi_3+\phi_4) \right)  \rangle  ^\mathrm{conn.} ~,  \label{Eq:c4ani_b}
\eea
which determines the azimuthal anisotropy from four particle
correlations: $v_2\{4\} = (-c_2\{4\})^{1/4}$. Before addressing the
corrections written in Eqs.~(\ref{Eq:c4ani_a},\ref{Eq:c4ani_b}) we compute the
fully connected contribution and show that it is positive.

The fully connected contribution to the S-matrix is given by
\be
(N_c^2-1) \, \prod_{i=1}^{4}
\frac{-Q_s^2}{4(N_c^2-1)}
(\vec r_i\cdot \vec r_{i+1})\,
\Delta(\vec b_i-\vec b_{i+1})\, \log\frac{1}{r_i\Lambda}
\;+\; \mathrm{permutations}~,
\ee
where $i+1$ is defined modulo $4$. Averaging over impact parameters
generates a factor of $1/(4 N_D^{3})$. We may now perform the Fourier
transform and sum the 48 contractions of
the amplitudes / conjugate amplitudes of dipoles 1 to 4. This leads
to
\be
\langle \exp\left( 2i (\phi_1 +\phi_2 - \phi_3
-\phi_4)\right)  \rangle^\mathrm{conn.} =
\frac{1}{4 N_D^3} \frac{1}{(N_c^2-1)^3} (1+8{\cal A}^2)~.
\ee
Here, corrections of order $\sim 1/(N_c^2-1)$ to the normalization
have been neglected, see related discussion for $v_2\{2\}$ above. As
promised, the fully connected contribution to $c_2\{4\}$ is positive;
thus if the anisotropy ${\cal A}$ is zero, the elliptic harmonic
$v_2\{4\}$ would be {\it complex}.  Furthermore, the magnitude of the
fully connected contribution relative to $v_2\{1\}^4$ is $\sim
1/({\cal A}^4 N_c^6)$. Hence, parametrically $c_2\{4\}$  crosses
zero when ${\cal A}\sim 1/N_c^{3/2}$.

The terms from Eqs.~(\ref{Eq:c4ani_a},\ref{Eq:c4ani_b}), to leading
order in $N_c$, are given by
\bea
& &
\frac{1}{N_D^2} (v_2\{1\})^2 \left<\exp\; -2i(\phi_3+\phi_4)\right>^\mathrm{conn.}
= \frac{1}{N_D^3} \frac{ {\cal A}^4  }{N_c^2-1}\,,  \\
& &
\frac{1}{N_D}
\left<\exp\; 2i(\phi_1+\phi_2)\right>^\mathrm{conn.}\;
\left<\exp\; -2i(\phi_3+\phi_4)\right>^\mathrm{conn.}
= \frac{1}{N_D^3} \frac{ {\cal A}^4  }{(N_c^2-1)^2}\,, \\
& &
\frac{4}{N_D} v_2\{1\} \left<\exp\; 2i(\phi_1+\phi_2-\phi_3)\right>^\mathrm{conn.}
= \frac{8}{3N_D^3} \frac{ {\cal A}^4  }{(N_c^2-1)^2} ~.
\eea
They provide manifestly positive contributions to $c_2\{4\}$. 
When ${\cal A}$ is of order of $N_c^{-3/2}$, which is the regime where
$c_2\{4\}$ changes sign, we can write our final result in the form
\be \label{eq:c2_4_final} c_2\{4\} = - \frac{1}{N_D^3} \left( {\cal
  A}^4 - \frac{1}{4(N_c^2-1)^3} \right)~, 
\ee
Here the additional terms listed in
Eqs.~(\ref{Eq:c4ani_a},\ref{Eq:c4ani_b}) are suppressed by additional
powers of $N_c^{-2}$.

\section{Discussion}

An anisotropic single-particle distribution, $v_2\{1\}\ne0$, requires
an angular dependence of the dipole S-matrix $\sim \mathrm{tr}\,(\vec
r\cdot\vec E)^2$ for individual configurations of $\vec E$. We
describe this by the term $\sim {\cal A} (\hat r \cdot \hat a)^2$ in
Eq.~(\ref{eq:Ei_Ej_a}).

Our main results are as follows. The two-particle elliptic asymmetry
$c_2\{2\} \equiv (v_2\{2\})^2$ is given by
\be \label{eq:c2_2_total}
c_2\{2\} = \frac{1}{N_D}\;\left({\cal A}^2 + \frac{1}{4(N_c^2-1)}\right)
= \frac{1}{N_D}\;\left((v_2\{1\})^2 + \frac{1}{4(N_c^2-1)}\right)~.
\ee
The first term corresponds to the square of the asymmetry of the
one-particle distribution while the second term is due to
non-factorizable, genuine two-particle correlations.  The transition
between the two regimes occurs at ${\cal A}\sim 1/N_c$. In practice,
using $N_c=3$ and the estimate ${\cal A}\simeq0.2$ from
Ref.~\cite{Dumitru:2014dra} we conclude that the magnitudes of both
terms are comparable.

The elliptic asymmetry from four-particle correlations, $c_2\{4\}
\equiv - (v_2\{4\})^4$, is
\be \label{eq:c2_4_total}
c_2\{4\} = - \frac{1}{N_D^3}
\left[ (v_2\{1\})^4 
- \frac{1}{4(N_c^2-1)^3} \right] ~.
\ee
This expression applies when $v_2\{1\}={\cal O}(N_c^{-3/2})$, where
$c_2\{4\}$ changes sign. The first term on the r.h.s.\ corresponds to
the fully factorized distribution and is the only negative
contribution to $c_2\{4\}$. Thus, parametrically this transition to
$c_2\{4\}<0$ occurs {\em before} the one-particle factorizable
contribution dominates $c_2\{2\}$. That is, in the vicinity of
$c_2\{4\}=0$ the two-particle cumulant $c_2\{2\}$ is dominated at
leading order in $1/N_c^2$ by connected diagrams. We repeat, also, that
{\em all} contributions in
eqs.~(\ref{eq:c2_2_total},\ref{eq:c2_4_total}) computed within
small-$x$ QCD are long range in rapidity.

Our analysis naturally raises a question about the magnitude of the
$\vec E$-field polarization amplitude ${\cal A}$ and its dependence on
multiplicity. Averaging over all target configurations without a
multiplicity bias gives ${\cal A}\sim 0.1-0.15$ at small
$x$~\cite{aniso_MV_JIMWLK}. In fact, ${\cal A}(r)$ exhibits a (weak)
dependence on $r$ at small $r$ and this function has been
found~\cite{aniso_MV_JIMWLK} to coincide with the distribution of
linearly polarized gluons (for the MV model) obtained in
refs.~\cite{TMD}. The effect of a multiplicity bias remains to be
investigated. In order for the disconnected contribution to dominate
in high multiplicity events, ${\cal A}$ would have to grow with
multiplicity.

Although our present discussion is restricted to high-$p_\perp$
particles, i.e.\ small dipoles, it suggests that the measurement by
the ALICE and CMS collaborations of a sign change of $c_2\{4\}$
corresponds to the fully factorizable contribution becoming
dominant. The emergence of ``collectivity'' in pA collisions could be
viewed as multi-particle correlation functions becoming dominated by
fully disconnected diagrams, analogous to the BBGKY hierarchy. It will
be important to understand specifically how this emerges from
small-$x$ QCD dynamics.

\begin{acknowledgments}
We are grateful to Ho-Ung Yee for his contribution during 
the early stage of the project.
A.D.\ and V.S.\ thank G.~Denicol, Yu.~Kovchegov, S.~Schlichting,
P.~Sorensen and R.~Venugopalan for useful
discussions. A.D.\ gratefully acknowledges support by the DOE Office
of Nuclear Physics through Grant No.\ DE-FG02-09ER41620 and from The
City University of New York through the PSC-CUNY Research Award
Program, grant 67119-0045.
\end{acknowledgments}


\end{document}